\documentclass[traditabstract]{aa}
\usepackage{graphicx}
\usepackage{natbib}
\usepackage{txfonts}
\usepackage{ulem}
\usepackage{marginnote}
\usepackage{subfigure}
      \def\new#1 {{\bf #1 }}
      \def\cut#1 {\sout{#1} }

\def\sol {$_{\odot}$}

\def\percc {$\mathrm{cm^{-3}}$} 
\def\cmsq  {$\hbox{{\rm cm}}^{-2}$}    

\def\CSEO {\hbox{${\rm C}^{17}{\rm O}$}}   
\def\AMM {\hbox{${\rm NH}_{3}$}} 

\def\HCOP {HCO$^+$}

\def\HII{H{\sc ii}}

\def\folio{\ifnum\pageno=1\nopagenumbers\else\number\pageno\fi}

\def\lax    {\ifmmode{_<\atop^{\sim}}\else{${_<\atop^{\sim}}$}\fi}
\def\gax    {\ifmmode{_>\atop^{\sim}}\else{${_>\atop^{\sim}}$}\fi}
\newbox\grsign      \setbox\grsign=\hbox{$>$} 
\newdimen\grdimen   \grdimen=\ht\grsign
\newbox\simgreatbox \setbox\simgreatbox=\hbox{\raise.5ex\hbox{$>$}\llap
                        {\lower.5ex\hbox{$\sim$}}}\ht1=\grdimen\dp1=0pt
\newbox\simlessbox  \setbox\simlessbox =\hbox{\raise.5ex\hbox{$<$}\llap
                        {\lower.5ex\hbox{$\sim$}}}\ht2=\grdimen\dp2=0pt

\newbox\grsign \setbox\grsign=\hbox{$>$} \newdimen\grdimen \grdimen=\ht\grsign
\newbox\laxbox \newbox\gaxbox
\setbox\gaxbox=\hbox{\raise.5ex\hbox{$>$}\llap
     {\lower.5ex\hbox{$\sim$}}}\ht1=\grdimen\dp1=0pt
\setbox\laxbox=\hbox{\raise.5ex\hbox{$<$}\llap
     {\lower.5ex\hbox{$\sim$}}}\ht2=\grdimen\dp2=0pt
\def\gax{\mathrel{\copy\gaxbox}}
\def\lax{\mathrel{\copy\laxbox}}
\def\boxit#1    {\vbox{\hrule\hbox{\vrule\kern3pt
                  \vbox{\kern3pt#1\kern3pt}\kern3pt\vrule}\hrule}}
\def\h      {\ifmmode{^{\rm h}}\else{$^{\rm h}$}\fi}
\def\m      {\ifmmode{^{\rm m}}\else{$^{\rm m}$}\fi}
\def\s      {\ifmmode{^{\rm s}}\else{$^{\rm s}$}\fi}
\def\decas    {\ifmmode{{\rlap.}{''}}\else{${\rlap.}{''}$}\fi}
\def\mum     {\ifmmode{\mu{\rm m}}\else{$\mu{\rm m}$}\fi}
\def\s      {\ifmmode{^{\rm s}}\else{$^{\rm s}$}\fi}
\def\deg      {\ifmmode{^{\circ}}\else{$^{\circ}$}\fi}
\def\as     {\ifmmode {\rlap.}$\,$''$\,$\! \else ${\rlap.}$\,$''$\,$\!$\fi}
\def\decsec  {\ifmmode {\rlap.}$\,$^{s}$\,$\! \else ${\rlap.}$\,$^{s}$\,$\!$\fi}\def\decs  {\ifmmode {\rlap.}$\,$^{s}$\,$\! \else ${\rlap.}$\,$^{s}$\,$\!$\fi}

\def\kms    {\ifmmode{{\rm km~s}^{-1}}\else{km~s$^{-1}$}\fi}

\def\Mspy   {\ifmmode {M_{\odot} {\rm yr}^{-1}} \else $M_{\odot}$~yr$^{-1}$\fi}
\def\Mdot   {\ifmmode {\dot M} \else $\dot M$\fi}
\def\mhd    {\ifmmode {n_{{\rm H}_2}} \else $n_{{\rm H}_2}$\fi}
\def\mhcd   {\ifmmode {N_{{\rm H}_2}} \else $N_{{\rm H}_2}$\fi}

\def\El      {\ifmmode{E_{\ell}}\else{$E_{\ell}$}\fi}
\def\beam    {\ifmmode{\theta_{\rm B}}\else{$\theta_{\rm B}$}\fi}
\def\mjyb   {\ifmmode {{\rm mJy~beam}^{-1}} \else{mJy~beam$^{-1}$}\fi}
\def\mujyb   {\ifmmode {\mu{\rm Jy~beam}^{-1}} \else{$\mu$Jy~beam$^{-1}$}\fi}
\def\Trot   {\ifmmode{T_{\rm rot}}\else$T_{\rm rot}$\fi}    
    
\def\Teff   {\ifmmode{T_{\rm eff}}\else$T_{\rm eff}$\fi}

\def\ITRS   {\ifmmode{\smallint {\rm T}_{R}^{*}dv}\else{$\smallint 
{\rm T}_{R}^{*}dv$}\fi}
\def\ITRS   {\ifmmode{\smallint {\rm T}_{R}^{*}dv}\else{$\smallint 
{\rm T}_{R}^{*}dv$}\fi}
\def\ITAS   {\ifmmode{\smallint {\rm T}_{A}^{*}dv}\else{$\smallint 
{\rm T}_{A}^{*}dv$}\fi}

\def\HII        {H~{\eightpt II}}

\def\lefttitle#1  {\noindent \hangindent=18.0pt \hangafter=1 {#1} \par}
\def\vol#1  {{\bf {#1}{\rm,}\ }}
\font\eightpt=cmr8

\font\tenssb=cmssbx10
\font\tenbf=cmbx10
\font\sevenbf=cmbx8
\font\fivebf=cmbx6
\def\unetdemi    {\smallskipamount=6pt plus2pt minus2pt
                  \medskipamount=12pt plus4pt minus4pt
                  \bigskipamount=24pt plus8pt minus8pt
                  \normalbaselineskip=16pt plus0pt minus0pt
                  \normallineskip=2pt
                  \normallineskiplimit=0pt
                  \jot=6pt
                  {\def\smallskip {\vskip\smallskipamount}}
                  {\def\medskip   {\vskip\medskipamount}}
                  {\def\bigskip   {\vskip\bigskipamount}}
                  {\setbox\strutbox=\hbox{\vrule 
                    height17.0pt depth7.0pt width 0pt}}
                  \parskip 12.0pt
                  \normalbaselines}
\def\smallerspace {\smallskipamount=3pt plus0pt minus0pt
                  \medskipamount=6pt plus0pt minus0pt
                  \bigskipamount=10.5pt plus0pt minus0pt
                  \normalbaselineskip=10.5pt plus0pt minus0pt
                  \normallineskip=1pt
                  \normallineskiplimit=0pt
                  \jot=3pt
                  {\def\smallskip {\vskip\smallskipamount}}
                  {\def\medskip   {\vskip\medskipamount}}
                  {\def\bigskip   {\vskip\bigskipamount}}
                  {\setbox\strutbox=\hbox{\vrule 
                    height8.5pt depth3.5pt width 0pt}}
                  \parskip 0pt
                  \normalbaselines}
\def\memospace    {\smallskipamount=4pt plus1pt minus1pt
                  \medskipamount=6pt plus2pt minus2pt
                  \bigskipamount=14pt plus6pt minus6pt
                  \normalbaselineskip=14pt plus0pt minus0pt
                  \normallineskip=1pt
                  \normallineskiplimit=0pt
                  \jot=4pt
                  {\def\smallskip {\vskip\smallskipamount}}
                  {\def\medskip   {\vskip\medskipamount}}
                  {\def\bigskip   {\vskip\bigskipamount}}
                  {\setbox\strutbox=\hbox{\vrule 
                    height17.0pt depth7.0pt width 0pt}}
                  \parskip 2.0pt
                  \normalbaselines}
\def\memowidespace    {\smallskipamount=5pt plus1pt minus1pt
                  \medskipamount=7.5pt plus2pt minus2pt
                  \bigskipamount=17.5pt plus6pt minus6pt
                  \normalbaselineskip=17.0pt plus0pt minus0pt
                  \normallineskip=1.25pt
                  \normallineskiplimit=0pt
                  \jot=5pt
                  {\def\smallskip {\vskip\smallskipamount}}
                  {\def\medskip   {\vskip\medskipamount}}
                  {\def\bigskip   {\vskip\bigskipamount}}
                  {\setbox\strutbox=\hbox{\vrule 
                    height21.25pt depth8.75pt width 0pt}}
                  \parskip 2.5pt
                  \normalbaselines}
\usepackage{color}

\begin{document}

\title{Infall through the evolution of high-mass star-forming clumps} 

\author{F. Wyrowski \inst{1} \and R. G\"usten \inst{1} \and
  K. M. Menten \inst{1} \and H. Wiesemeyer \inst{1}  \and T. Csengeri \inst{1}
  \and S. Heyminck \inst{1} \and B. Klein \inst{1,2} \and C. K\"onig \inst{1}
  \and J.S. Urquhart \inst{1}}

\offprints{F. Wyrowski}

\institute{Max-Planck-Institut f\"ur Radioastronomie,
Auf dem H\"ugel 69, D-53121 Bonn, Germany \and
University of Applied Sciences Bonn-Rhein-Sieg, Grantham-Allee 20,
53757, Sankt Augustin, Germany
\\
\email{wyrowski, rguesten, kmenten, hwiese, csengeri, heyminck, bklein, ckoenig, jurquhart@mpifr-bonn.mpg.de}}

\date{Received / Accepted}


\authorrunning{Wyrowski et al.}

\abstract {
  With the GREAT receiver at the Stratospheric Observatory for
  Infrared Astronomy (SOFIA), nine massive molecular clumps have been
  observed in the ammonia $3_{2+}- 2_{2-}$ line at 1.8~THz in
a search
  for signatures of infall.  The sources were selected from the
  ATLASGAL submillimeter dust continuum survey of our Galaxy. Clumps
  with high masses covering a range of evolutionary stages based on
  their infrared properties were chosen. The ammonia line was detected
  in all sources, leading to five new detections and one confirmation of a
    previous detection of redshifted absorption in front of their
  strong THz continuum as a probe of infall in the clumps. These
  detections include two clumps embedded in infrared dark clouds. The
  measured velocity shifts of the absorptions compared to optically
  thin \CSEO\ (3--2) emission are 0.3--2.8~km/s, corresponding to fractions of 3\%\ to
  30\% of the free-fall velocities of the clumps.  The
  ammonia infall signature is compared with complementary data of
  different transitions of HCN, HNC, CS, and HCO$^+$, which are often
  used to probe infall via their blue-skewed line profiles. The
  best agreement with the ammonia results is found for the HCO$^+$
  (4--3) transitions, but the latter is still strongly blended with
  emission from associated outflows. This outflow signature is far
  less prominent in the THz ammonia lines, which confirms it as a powerful
  probe of infall in molecular clumps.  Infall rates in the range from
  0.3 to 16~$10^{-3}\,M_\odot/$yr were derived with a tentative
  correlation with the virial parameters of the clumps. The new
  observations show that infall on clump scales is ubiquitous through
  a wide range of evolutionary stages, from $L/M$ covering about ten
  to several hundreds.
}


\maketitle

\section{\label{intro}Introduction}

Observation of infall is key to our understanding of the accretion
process in star formation. High-resolution spectroscopy
allows us to resolve molecular lines originating from the dense molecular
envelopes of the forming (proto-) stars to deduce the kinematics of the
gas.  Optically thin emission lines probe weighted mean velocities
through the whole envelope, hence they can be used to measure the
systemic velocities of the clumps in which the forming stars
reside. Optically thick lines are very sensitive to excitation
gradients within the clumps.  Such a gradient leads to differences
in the blue- and redshifted peak temperatures of self-absorbed
spectra: in case of infall and rising excitation toward the inner
part of the clump, high optical depths will lead to probing the outer envelope at lower excitation with the
redshifted gas, whereas the
blueshifted gas is seen preferentially from the inner envelope at
higher excitation and hence peak temperature. This process will lead
to the well-known blue-skewed line profiles as a signature of
infall. An excellent review of this method, including its many pitfalls,
is given in \citet{evans2002}. In addition to the excitation gradient, the
optical depths need to attain proper values to allow for the
separation of the two line peaks. Furthermore, any outflow activity in
the clump can easily blend with the infall signature and obscure it.

To avoid these complications, redshifted absorption in front of bright
continuum sources can be used to probe the velocity field on the
line of sight toward the continuum. As continuum can serve either the
centimeter free-free emission of \HII\ regions
\citep{zhang+1997, sollins+2005, beltran+2006} or -- as recently shown
by \citet{wyrowski+2012} -- the bright submillimeter dust continuum
emission of the clumps itself. In addition to the continuum background, this
method requires transitions with high critical density so that the
excitation temperature in the envelope is lower than the temperature
of the continuum background. Given the typical high Einstein $A$
values of hydride lines in the THz range, this can easily be
achieved either with ground-state lines or lines originating from
metastable levels such as the ($J$, $K$) states with $J=K$ in
ammonia. Compared to absorption studies toward \HII\ regions,
the latter method can be employed toward a much wider range of
evolutionary stages of massive clumps.

With the aim of studying infall through the different evolutionary
stages of massive star-forming clumps, the selection of targets
benefits enormously from unbiased Galactic plane surveys that have
been conducted in the past years at mid-infrared
\citep[GLIMPSE/MIPSGAL/WISE]{churchwell+2009,carey+2009,wright+2010}
and far-infrared or submillimeter wavelengths \citep[Hi-GAL;
  ATLASGAL; BGPS]{molinari+2010b,schuller+2009, ginsburg+2013}. In
particular the ATLASGAL survey at 870~$\mu$m in combination with the
MIR surveys is well suited to select the most massive clumps in a
range of evolutionary stages owing to its relatively high angular
resolution at a long wavelength where the dust emission is still
optically thin.

\citet{wyrowski+2012} presented observations toward three massive clumps
with the Stratospheric Observatory for Infrared Astronomy (SOFIA) in
the ammonia $J_K=3_{2+} - 2_{2-}$  at 1.8~THz to
measure the infall signatures of the clumps. Ammonia absorption was detected in
all of them, showing clear evidence for infall in each case with
infall rates between 3 and $10\times 10^{-3}~M$\sol/yr. The spectra
were modeled with the 1D radiative transfer code RATRAN
\citep{hogerheijde+2000} using the physical structure of
\citet{rolffs+2011} and a variety of velocity fields.

 \begin{table*}[th]
  \caption{Ammonia source sample and observing parameters. Distances,
   luminosities, and masses are taken from K\"onig et al. (subm.) except for G5.89 
   and W33A, which are taken from \citet{vdtak+2013}. Virial parameters are adopted from \citet{giannetti+2014}.
   Additionally, updated parameters for sources from \citet{wyrowski+2012} are listed.}
  \label{t:sample}
  \begin{center}
  \begin{tabular}{llrrcccccc}
  \hline
  \hline
  Source & Stage       & R.A.(J2000)    & Decl.(J2000)  &  $L_{\rm bol}$  & $M$ & $\alpha_{\rm
  vir}$ & $d$  & $T_{\rm sys}$ & $t_{ON}$   \\
        &        & ( h:m:s ) & ( $^\circ$:$^\prime$:$^{\prime\prime}$ )      &  ($L_\odot$)  & ($M_\odot$) &  & (pc) & (K) & (min.)  \\
\hline 
G34.41+0.2    & IRDC &  18:53:18.13  &   01:25:23.7  & 4.8(3) & 1.5(2)& 1.68 & 1600 & 1990 & 11.9 \\
G23.21-0.3    & IRDC &  18:34:54.91  & --08:49:19.2  & 1.3(4) & 1.1(3)& 0.57 & 4600 & 1600 & 23.8 \\
G327.29-0.6   & HC   &  15:53:07.80  & --54:37:06.4  & 8.2(4) & 2.4(3)& 1.72 & 3100 & 1960 &  9.4 \\
G34.26+0.2    & UCHII&  18:53:18.49  &   01:14:58.7  & 4.6(4) & 6.8(2)& 0.95 & 1600 & 1920 & 18.8 \\
G351.58-0.4   & UCHII&  17:25:25.03  & --36:12:45.3  & 2.4(5) & 7.4(3)& 0.42 & 6800 & 2020 & 18.8 \\
G35.20-0.7    & MYSO &  18:58:12.93  &   01:40:40.6  & 2.5(4) & 3.8(2)& 1.11 & 2200 & 1940 & 18.2 \\
G5.89-0.4     & UCHII&  18:00:30.40  &  -24:04:00.0  & 3.6(4) & 1.4(2)&      & 1300 & 1890 & 12.5 \\
W33A          & MYSO &  18:14:39.52  & --17:52:00.5  & 3.0(4) & 7.0(2)&      & 2400 & 2000 &  6.2 \\
W49N          & UCHII&  19:10:13.20  &   09:06:12.0  & 3.6(6) & 3.4(4)&      &11100 & 1930 & 10.7 \\
\hline
W43MM1        & MYSO &  \multicolumn{2}{c}{\citet{wyrowski+2012}} & 6.3(4) & 4.6(3)& 1.12 & 4900 & --- & --- \\
G31.41+0.31   & HC   &  \multicolumn{2}{c}{\citet{wyrowski+2012}} & 6.7(4) & 2.7(3)& 0.61 & 4900 & --- & --- \\
\hline
 \end{tabular}
 \end{center}
\end{table*}

Taking advantage of the unbiased nature of the ATLASGAL dust continuum
survey of the inner Galaxy, we selected new flux-limited samples of
ATLASGAL sources with additional infrared selection criteria that
ensure that we cover a broader range of evolutionary stages and luminosity.
The systemic velocities of all of them have recently been measured
using C$^{17}$O (3--2) \citep{giannetti+2014}. The
whole sample was characterized using the ATLASGAL data in
combination with \textit{Herschel} (K\"onig et al., subm.) and ground-based
molecular line follow-up programs to allow detailed modeling of the
clumps. Here, we present an extension of our SOFIA ammonia observing
program to additional stages of massive star-forming clumps selected from
the samples above, including earlier stages within infrared dark clouds.

\section{\label{obs}Observations and data reduction}

\subsection{SOFIA}

We used the L2 channel of the GREAT\footnote{GREAT is a development by
  the MPI f\"ur Radioastronomie and the KOSMA/Universit\"at zu K\"oln,
  in cooperation with the MPI f\"ur Sonnensystemforschung and the DLR
  Institut f\"ur Planetenforschung.} instrument \citep{heyminck+2012}
onboard SOFIA during Cycle 1 to observe the NH$_3$ $3_{2+}
- 2_{2-}$ line at 1810.379~GHz. In contrast to the situation
  reported by \citet{wyrowski+2012}, we were able to tune the line into the
upper sideband of the receiver to avoid a deep telluric absorption at
1812.5~GHz.
As backends, fast Fourier transform spectrometers
\citep{Klein+2012} were used (XFFTS) to cover a bandwidth of
2.5~GHz. The velocity resolution was smoothed to 0.5~km/s to increase
the signal-to-noise ratio in the spectra.  The beam size at the
observing frequency is 16 arcsec.

The positions of the observed sources are given in
Table~\ref{t:sample}. G34.26+0.15, which has been presented
   by \citet{wyrowski+2012}, was reobserved to examine the
line profile with a higher signal-to-noise.   The nominal focus position
was updated regularly against temperature drifts of the telescope
structure. The pointing was established with the optical guide cameras
to an accuracy of a few arcsec. All of the data were taken during the
southern deployment flights from New Zealand in July 2013 under
excellent observing conditions, resulting in SSB system temperatures at
the ammonia frequency ranging from 1450 to 2050~K.

Spectra were calibrated to a $T_{\rm A}^*$ scale and were then
converted to $T_{\rm MB}$  assuming a forward efficiency of
0.97 and beam efficiencies of 0.67.  Final processing and analysis of
the data was then performed using the CLASS program within the GILDAS
software \footnote{http://www.iram.fr/IRAMFR/GILDAS}.

\subsubsection{Dust continuum levels}
 
\begin{figure}[t]
\begin{center}
\includegraphics[width=0.45\textwidth,angle=0]{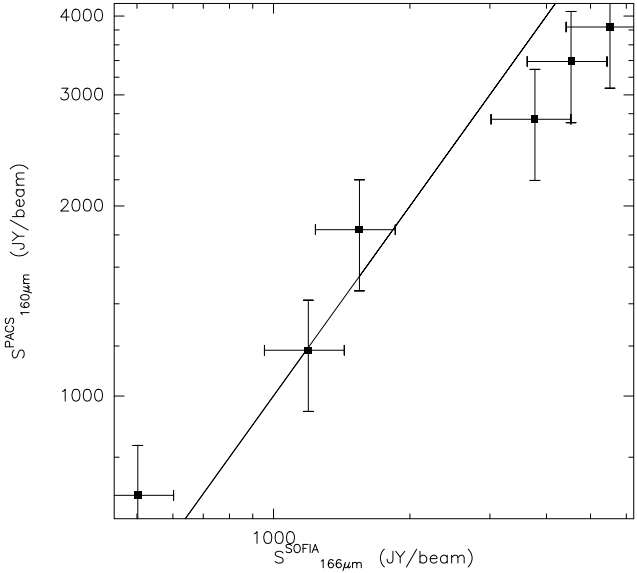}
\caption{\label{fig:cont} Comparison of GREAT continuum levels with
  PACS 160~$\mum$ flux densities with nominal 20\% errors. To account
  for the different beam sizes, we scaled the PACS fluxes with the
  ratio of the beam areas since the typical FWHM sizes from PACS are
  about 20''. The straight line indicates equal flux densities. }
\end{center}
\end{figure}

To obtain reliable detections of the continuum by chopping off the
atmospheric total power, the wobbling secondary was used with a
typical throw of $120''$ about the cross elevation axis in a symmetric
mode.
Figure~\ref{fig:cont} shows the comparison with the PACS 160~$\mu$m
continuum extracted from the Herschel Hi-GAL data (K\"onig et al.,
subm.) and scaled with the areas of the different beams. Given the
differences in observing beams, the fluxes agree reasonably well.  

\subsection{Ground-based complementary data}

In addition, ground-based spectra of dense gas tracers were used to
complement the interpretation of the SOFIA ammonia absorption
observations. Table~\ref{tab:ground} gives an overview of the lines.
The systemic velocities from C$^{17}$O (3--2) were taken from
\citet{giannetti+2014}, who also list details about the IRAM 30m
  \footnote{Based on observations carried out with the IRAM 30m telescope. IRAM
  is supported by INSU/CNRS (France), MPG (Germany) and IGN (Spain).}
 and APEX 
  \footnote{Based on observations with the Atacama Pathfinder EXperiment (APEX)
  telescope. APEX is a collaboration between the Max Planck Institute
  for Radio Astronomy, the European Southern Observatory, and the
  Onsala Space Observatory.}
observations. Details about the Mopra observations of dense
  gas tracers will be presented in Wyrowski et al. (in prep.).

\begin{table}
  \caption{Complementary ground-based observations of dense gas tracers}
\label{tab:ground} 
\begin{center}
 \begin{tabular}{lcrl}
 \hline\hline 
Molecule & Transition  & Frequency (MHz) & Telescopes \\
\hline 
CS       &  2--1   &  97980.950 &  IRAM 30m \\
         &  7--6   & 342883.000 &  APEX \\
HCO$^+$  &  1--0   &  89188.523 &  IRAM 30m/Mopra \\
         &  4--3   & 356734.134 &  APEX \\
HNC      &  1--0   &  90663.593 &  IRAM 30m/Mopra \\ 
         &  4--3   & 362630.303 &  APEX \\
HCN      &  4--3   & 354505.476 &  APEX \\
\hline
 \end{tabular}
 \end{center}
\end{table}

\section{Results}
\label{sec:results}

\subsection{Ammonia absorption}

%
%
%

\begin{figure*}[t]
\begin{center}
\includegraphics[width=0.85\textwidth,angle=0]{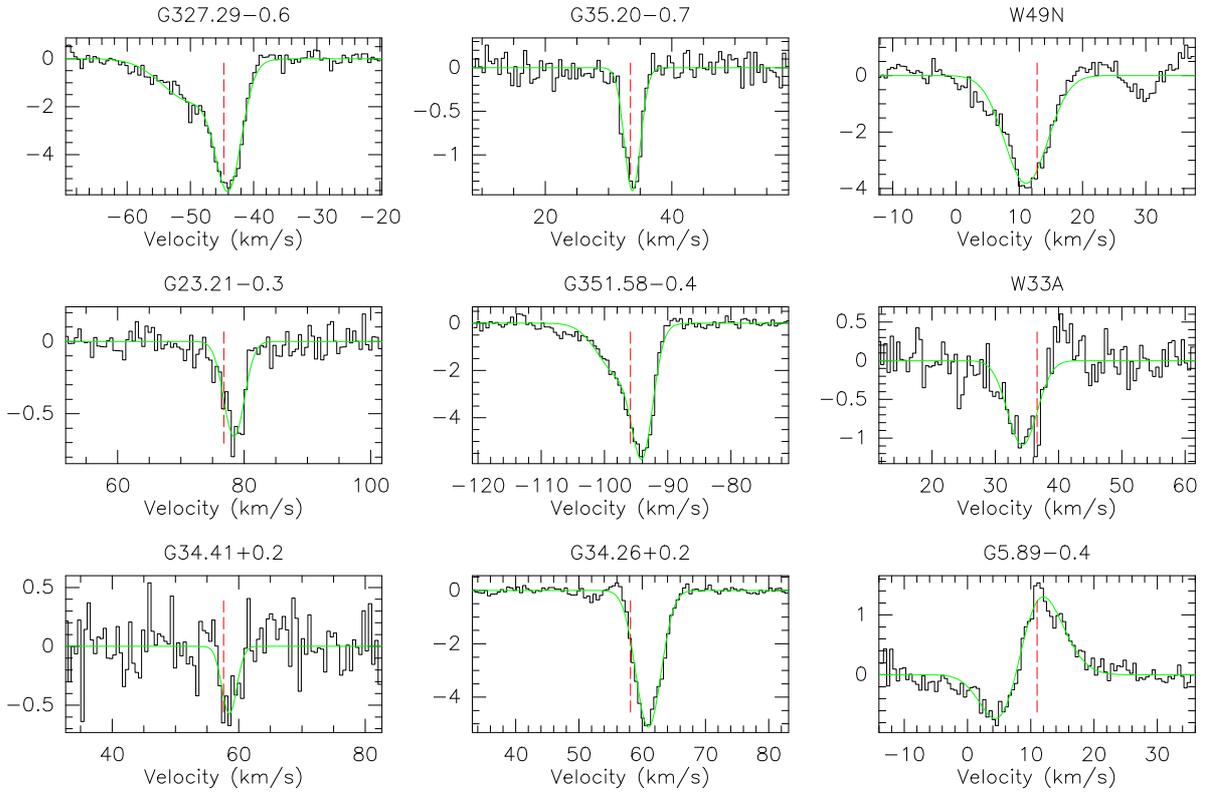}
\caption{\label{fig:spectra} \AMM\ $3_{2+} - 2_{2-}$ spectra of the
  observed sources.  Results of Gaussian fits to the line profiles are overlaid
  in green. The systemic velocities of the sources, determined using
  C$^{17}$O (3--2), are shown with dotted lines. W49N in addition
  shows at 30m~km/s \AMM\ $3_{1+} - 2_{1-}$ from the other sideband. }
\end{center}
\end{figure*}

The averaged and baseline-subtracted spectra are shown on a $T_{\rm
  MB}$ scale in Fig.~\ref{fig:spectra}. Ammonia absorption lines were
detected in all sources, while G5.89-0.4 shows an additional
redshifted emission component. When we applied single Gaussian fits to
the line profiles, it became clear that several sources showed an
additional broader component. For G327.3--0.6, G351.58--0.3, and
G5.89-0.4 we therefore performed fits with an additional broad
Gaussian component to account for the high-velocity absorption
(emission in the case of G5.89-0.4). These fits are shown in
Fig.~\ref{fig:spectra} and are compared with C$^{17}$O (3--2) line
velocities. The APEX C$^{17}$O observations have been obtained
with a beam of 18~arcsec, very similar to the SOFIA beam at the
ammonia frequency, and they probe, as optically thin lines, the
systemic or averaged velocity of the massive clumps. 
 
While for our kinematical interpretation of the observed line profiles
the continuum is not relevant, a proper estimate of the continuum is
required to reliably determine the line-to-continuum ratio
that can be used to determine the opacity of the absorption lines and
in turn the column density in the absorbing level.  We therefore used
the line-to-continuum ratio to determine the opacities of the
absorption lines that are close to unity. They are in turn
proportional to the column densities in the lower level
\citep[see][]{wyrowski+2012}.

The corresponding
line and continuum parameters are given in Table~\ref{t:linepar}, which consist
  of the peak temperatures $T_{\rm MB}$, FWHM line widths $\Delta v$
  and velocities $v^{\rm NH_3}_{\rm LSR}$ and $v^{\rm C^{17}O}_{\rm
    LSR}$ of the Gaussian fits, the velocity differences $v_{\rm
    shift}$ beween the ammonia absorption and the C$^{17}$O emission
  lines, the measured continuum offsets $T_{\rm C}$ of the spectra and
  the line optical depths $\tau$ computed from the line-to-continuum
  ratios.
Five new clumps with redshifted absorption were found. For G34.26 the
redshifted absorption was reported in
\citet{wyrowski+2012}. G5.89-0.4 is dominated by ammonia from
outflowing gas (see, e.g., the recent SOFIA study of this source by
\citet{leurini+2015}, which also shows P Cygni profiles in water
  and OH, originating from the outflow). An expanding shell structure
  of W49N has been discussed in \citet{peng+2010} and most likely gives
  rise to the kinematics observed in the ammonia P-Cygni-like line
  profile. In W33A an outflow-dominated line profile similar to ammonia
  is detected in the\textit{ Herschel}/HIFI water ground-state line
  \citep{marseille+2010}.

\begin{table*}[t]
  \caption{Line parameters from Gaussian fits to the \AMM\ lines.
    Nominal fit errors are given in parenthesis. 
    $T_{\rm C}$ is the continuum level from the baseline fitting.
    In addition, the velocity of C$^{17}$O (3--2) lines observed with the APEX telescope are taken from \citet{giannetti+2014}. The optical depth is computed from the line-to-continuum ratio and subsequently used to compute the column density in the $2_{2-}$ level.
  }
\label{t:linepar} 
\begin{center}
 \begin{tabular}{lrrrrrrrc}
 \hline\hline 
Source & $T_{\rm MB}$  & $\Delta v$ & $v^{\rm NH_3}_{\rm LSR}$ & $v^{\rm C^{17}O}_{\rm LSR}$ & $v_{\rm shift}$ & $T_{\rm C}$ & $\tau$ & $N(2_{2-})$ \\
       & (K) & (\kms)    & (\kms) & (\kms) & (\kms) & (K) &  & ($10^{13}$\cmsq) \\
\hline 
G34.41+0.2 &  -0.57 (0.21)  &  2.80 (0.53) &  58.52 (0.31) &   57.63  & 0.89   & 1.76 & 0.4 & 1.2 \\
G23.21-0.3 &  -0.66 (0.07)  &  3.84 (0.31) &  78.40 (0.11) &   76.77  & 1.63   & 0.74 & 2.2 & 9.5 \\
G327.29-0.6&  -4.65 (0.21)  &  4.77 (0.16) & -43.92 (0.06) & -44.72   & 0.80    & 6.68  & 1.2 & 6.3 \\
           &  -1.79 (0.21)  & 11.11 (0.65) & -49.62 (0.44) &          &       &      & & \\
G34.26+0.2 &  -5.12 (0.08)  &  4.70 (0.05) &  60.97 (0.02) &   58.12  & 2.85   & 8.14 & 1.0 & 5.2 \\
G351.58-0.4&  -4.61 (0.15)  &  3.89 (0.15) & -93.95 (0.07) & -95.91   & 1.96   & 5.56  & 1.8 & 7.6 \\
           &  -2.14 (0.15)  &  7.69 (0.46) & -97.74 (0.35) &          &       &      & & \\
G35.20-0.7 &  -1.41 (0.12)  &  3.02 (0.16) &  33.76 (0.07) &   33.44  & 0.32   & 2.28 & 1.0 & 3.2 \\
 G5.89-0.4 &  -1.78 (0.16)  &  7.87 (0.93) &   6.32 (0.75) &   11.0   & -4.68  & 6.76 & 0.3 & 2.7 \\
           &   1.95 (0.16)  &  9.63 (0.66) &   9.93 (0.76) &          &       &      & & \\
      W33A &  -1.09 (0.20)  &  5.52 (0.50) &  34.29 (0.26) &   36.6   & -2.31  & 2.46 & 0.6 & 3.6 \\
      W49N &  -3.82 (0.43)  &  8.07 (0.37) &  11.09 (0.15) &   12.82  & -1.73  & 12.82& 0.4 & 3.2 \\
\hline
 \end{tabular}
 \end{center}
\end{table*}

\subsection{Dense gas probes}

\begin{figure}[ht]
\begin{center}
\includegraphics[height=0.73\textwidth,angle=0]{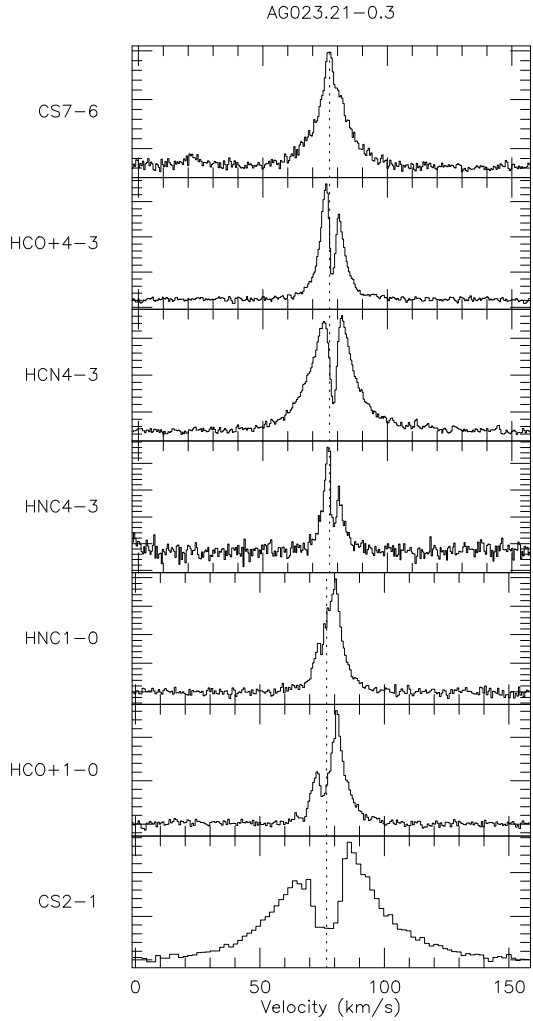}
\caption{\label{fig:densegas-g23} Ground-based observations of millimeter and submillimeter
transitions of the dense gas tracers HCN/HNC/CS/HCO$^+$ in G23.21--0.3 observed with
the IRAM 30m and APEX telescopes. The systemic velocity from \CSEO\ (3--2) is indicated with a dashed line. }
\end{center}
\end{figure}

  All of the ammonia absorption line sources presented here and in \citet{wyrowski+2012} were observed in dense gas
  tracers, with the exception of G5.89-0.4 and W33A.
Figures~\ref{fig:densegas-g23}, \ref{fig:densegas-1}, and
\ref{fig:densegas-2} show the observed line profiles. Using these
complementary data, we can compare the kinematical signatures of
different molecules with the observed ammonia absorption. In three
cases even lines from two different $J_{\rm upper}$ levels are
available. Clump G23.21-0.3, shown in Fig.~\ref{fig:densegas-g23}, is an
instructive example: while absorption redshifted by 1.6~km/s was
detected in ammonia, the line profiles from the dense gas probes do
not show consistent blue-skewed profiles. CS and HCN show broad line
wings indicative of a strong outflow that apparently in these lines
hides the signature of infalling gas in the envelope. HNC and HCO$^+$
show much weaker outflow wings, and their (4--3) transitions present
the blue-skewed signature of infall. Interestingly, the (1--0)
transitions of these molecules show the reverse. The critical density
of these lines differs by about a factor 50 (since the Einstein
$A_{ul}$ values scale with $\nu^3$), hence the (1--0) lines might be
dominated by low-density outer layers \citep{lopez-sepulcre+2010}.

Table~\ref{t:comparison} gives an overview of the derived kinematic
signatures from all the individual spectra we have available. While
there are clear variations for a given clump in different molecular
lines, the best agreement with the ammonia absorption results is found
for HCO$^+$ (4--3), and only W49N does not agree.

\begin{table*} 
   \caption{Comparison of kinematic signatures of ammonia absorption with ground-based
   observations of emission lines of HCN/HNC/CS/HCO$^+$. +/++ and -/-\,- correspond to weak
   and strong infall and expansion (or outflow) signatures, respectively. 0 indicate that
   no clear velocity trend is seen.
   }
   \label{t:comparison} 
   \begin{center}
     \begin{tabular}{lcccccccc}
       \hline\hline 
Source      & NH$_3$  & \multicolumn{2}{c}{HCO$^+$}  &  \multicolumn{2}{c}{HNC}   & \multicolumn{2}{c}{CS}         & HCN   \\
            &         & (1--0)   & (4--3) & (1--0) & (4--3) & (2--1) & (7--6) & (4--3) \\  
 \hline         
G327.29-0.6 &   +     &    0     &   0    &   0    &   0    &        &   +    &    0   \\
G351.58-0.4 &   ++    &    ++    &   +    &   ++   &   -    &        &   -    &    0   \\
G23.21-0.3  &   ++    &    -\,-    &   ++   &   -    &   ++   &  -\,-    &   0    &    0   \\
G34.41+0.2  &   +     &    -\,-    &   +    &   -\,-   &   0    &   0    &   0    &    +   \\
G35.20-0.7  &   +     &    ++    &   +    &   0    &   0    &  -\,-    &   0    &   -\,-   \\
G31.41+0.3  &   ++    &    -\,-    &   ++   &   -\,-   &  ++    &  ++    &   0    &   ++   \\
G34.26+0.2  &   ++    &    ++    &   ++   &   ++   &  ++    &  ++    &   +    &   ++   \\
G30.82-0.0  &   ++    &    +     &   ++   &   ++   &  ++    &        &   0    &   ++   \\
W49N        &   -     &    ++    &   +    &   +    &  +     &   0    &   +    &   ++   \\
 \hline
  \end{tabular}
  \end{center}
\end{table*}



\section{Analysis}
\label{s:analysis}

\subsection{Source sample}

\begin{figure}[ht]
\begin{center}
\includegraphics[width=0.45\textwidth,angle=0]{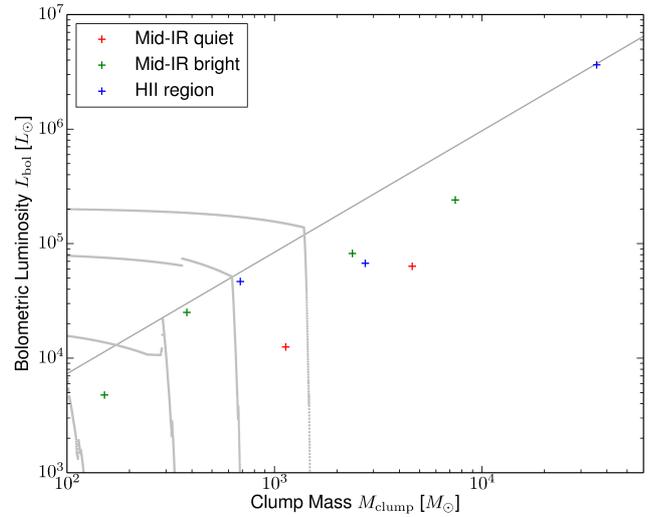}
\caption{\label{fig:sample} Mass-luminosity diagram of clumps with
  ammonia absorption line detections with evolutionary tracks as indicated in
  \citet{molinari+2008}: during the infall phase clumps
  move upward in luminosity, while later in the envelope clean-up
  phase, clumps move to the left when they lose their envelopes. }
\end{center}
\end{figure}

The sample of sources for which we have ammonia absorption lines
covers a range of luminosities and evolutionary phases.
Table~\ref{t:sample} gives the adopted distances and the luminosities
of the sources. Where applicable, we also added W43MM1 and G31.41 from \citet{wyrowski+2012} to our analysis here. The
luminosities were derived from fits to the spectral energy
distributions consisting of data from the Herschel Hi-GAL, WISE, and
ATLASGAL surveys, and they are discussed in detail in K\"onig et
al. (subm.). Figure~\ref{fig:sample} shows the corresponding
mass-luminosity diagram: our sample covers not only a wide range in clump masses, but also a
range of about 10 in luminosity-to-mass ratio, $L/M$, as an indicator
of evolutionary stage. Sources with even
lower $L/M$ will be difficult to observe in the 1.8~THz ammonia line because these clumps are colder and therefore become faint in the
continuum at 1.8~THz.

Different from the distance adopted by \citet{wyrowski+2012}, we here adopted  the
closer distance of K\"onig et al. (subm.) for G34.26+0.2. This
distance is based on recent
trigonometric parallax measurements toward neighboring clouds
(including G34.41 from our sample) \citep{kurayama+2011}.

\subsection{Physical structure of the clumps}

\begin{figure}[t]
\begin{center}
\includegraphics[width=0.45\textwidth,angle=0]{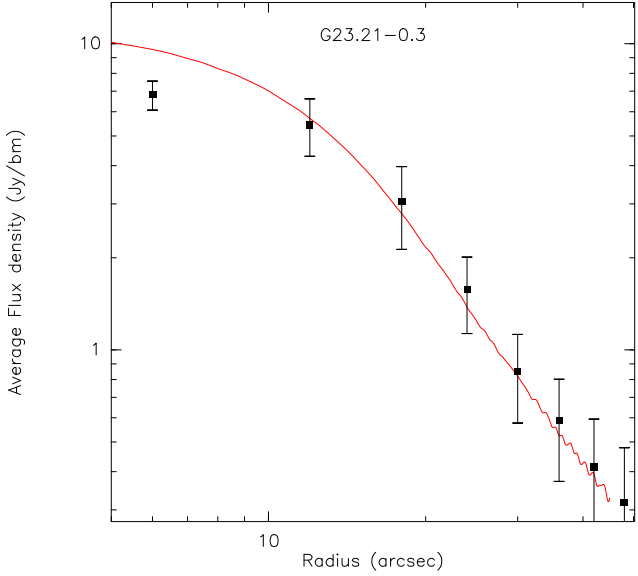}
\caption{\label{fig:ellint} Example of constraining the radial physical
                structure with the help of the 870~$\mu$m ATLASGAL submm dust continuum
                radial profiles in G23.21-0.3. As listed in Table~\ref{t:modelpar}, the
                fit corresponds to a $n\sim r^{-2}$ density power law and an outer radius of
                1.8~pc.}
\end{center}
\end{figure}

We constrained the physical structure of the clumps with luminosities
from the spectral energy distributions (Table~\ref{t:sample}) and the
submm dust continuum emission profiles from the ATLASGAL survey.  This
is similar to the approach in \citet{rolffs+2011}, whose results were
used for the modeling in \citet{wyrowski+2012}. The luminosities of
the clumps then dictate a dust temperature structure of the form
$T\sim L^{0.25}\times r^{-0.4}$
\citep[e.g.,][]{rowan-robinson1980}. The factor of proportionality was
adjusted to best fit the observed continuum at the ammonia frequency,
which led to agreement between model and observed continuum flux
within 20\% for a common factor.
The density
structure $n= n_{\rm 1pc}\times(r/1{\rm pc})^{\alpha_n}$ and outer radii
$R_{\rm out}$ of the clumps were constrained by computing the radial
structure of the submillimeter dust emission in circular annuli; an
example is given in Fig.~\ref{fig:ellint}. The inner radius was defined
by a dust destruction temperature of 1000~K, leading to radii of only
a few AU, so small compared to the rest of the envelope that the choice does
not influence the results of the modeling.

\subsection{Radiative transfer modeling}

\begin{table*}[th]
  \caption{Fit parameters from RATRAN modeling of the ammonia lines. The outer
    radius $R_{\rm out}$, density power law index $\alpha_n$ and density at 1~pc $n_{\rm 1pc}$
    were constrained with the 870~$\mu$m ATLASGAL dust continuum structure. 
    The kinematics is described by the turbulent widths $\delta v_t$ and the free-fall
    fraction $f_{ff}$. $X$(NH$_3$) and $X$(HCO$^+$) are best-fit abundances. 
  }
   \label{t:modelpar} 
   \begin{center}
     \begin{tabular}{lcccccccc}
       \hline\hline 
Source & $R_{\rm out}$  & $\alpha_n$ & $n_{\rm 1pc}$ & $\delta v_t$ & $f_{ff}$ & $X$(NH$_3$) & $X$(HCO$^+$) &  $\dot{M}$ \\
       & (pc)         &            & ($10^3$\percc)   & (km/s) &    & $10^{-8}$ & $10^{-10}$ & ($10^{-3}\,M_\odot/$yr) \\
 \hline         

G34.26+0.2  &  0.8  &  -1.7  &  10  &  2.4  &  0.3  &  0.19  &  0.25 &  9 \\
G327.29-0.6  &  2.  &  -1.9  &  10  &  2.3  &  0.05  &  0.5  &  0.2  &  4 \\
G351.58-0.4  &  1.8  &  -1.9 &  15  &  1.5  &  0.1  &  1.5  &  0.2  & 16 \\
G23.21-0.3  &  1.8  &  -2.0  &  4.5  &  1.0  &  0.2  &  1.5  &  0.5  &  8 \\
G35.20-0.7  &  1.5  &  -1.6  &  5.5  &  1.5  &  0.03  &  0.35&  0.3  &  0.3 \\
G34.41+0.2  &  1.0  &  -1.6  &  5  &  1.5  &  0.1  &  0.15   &  0.4  &  0.7 \\
 \hline         
  \end{tabular}
  \end{center}
\end{table*}

This physical structure of the envelopes was then implemented in RATRAN
models \citep{hogerheijde+2000} with power laws for the density and
temperature. We assumed the gas temperature to be equal to the dust
temperature.  The density structure was varied to reproduce the dust
continuum profiles.  This leaves the ammonia abundance $X$(\AMM), the velocity
fields, and the turbulence widths $\delta v_t$ as a function of radius as free parameters for the modeling
of the ammonia lines. Initial
guesses for the turbulence widths were taken from the observed \CSEO\
(3--2) line widths.

Similar to \citet{rolffs+2011} and \citet{wyrowski+2012}, the velocity
field was adjusted for sources showing redshifted absorption by
  varying $f_{ff}=v(r)/v_{ff}(r)$, hence the fraction of the free-fall
  velocity $v_{ff}=-\sqrt{2GM_{\rm enc}/r}$ caused at a given radius
  by the enclosed mass $M_{\rm enc}$. Figure~\ref{fig:hcop-nh3} shows the
results of the ammonia line modeling, and Table~\ref{t:modelpar} gives
the corresponding fit parameters. The fit parameters can be adjusted
to precisely fit the depth, widths, and red part of the absorption. But
several sources -- G23.21-0.3, G351.58-0.3, and G327.29-0.5 -- show
shallow blue absorption wings that are not reproduced by the fits because no
outflow component is included in the model. Nevertheless, the fit
methods allow us to reliably adjust the free-fall fraction $f_{ff}$
even in these cases.

\subsection{Simultaneous \HCOP\ modeling}

\begin{figure*}[t]
\begin{center}
\subfigure{
\includegraphics[width=0.35\textwidth,angle=0]{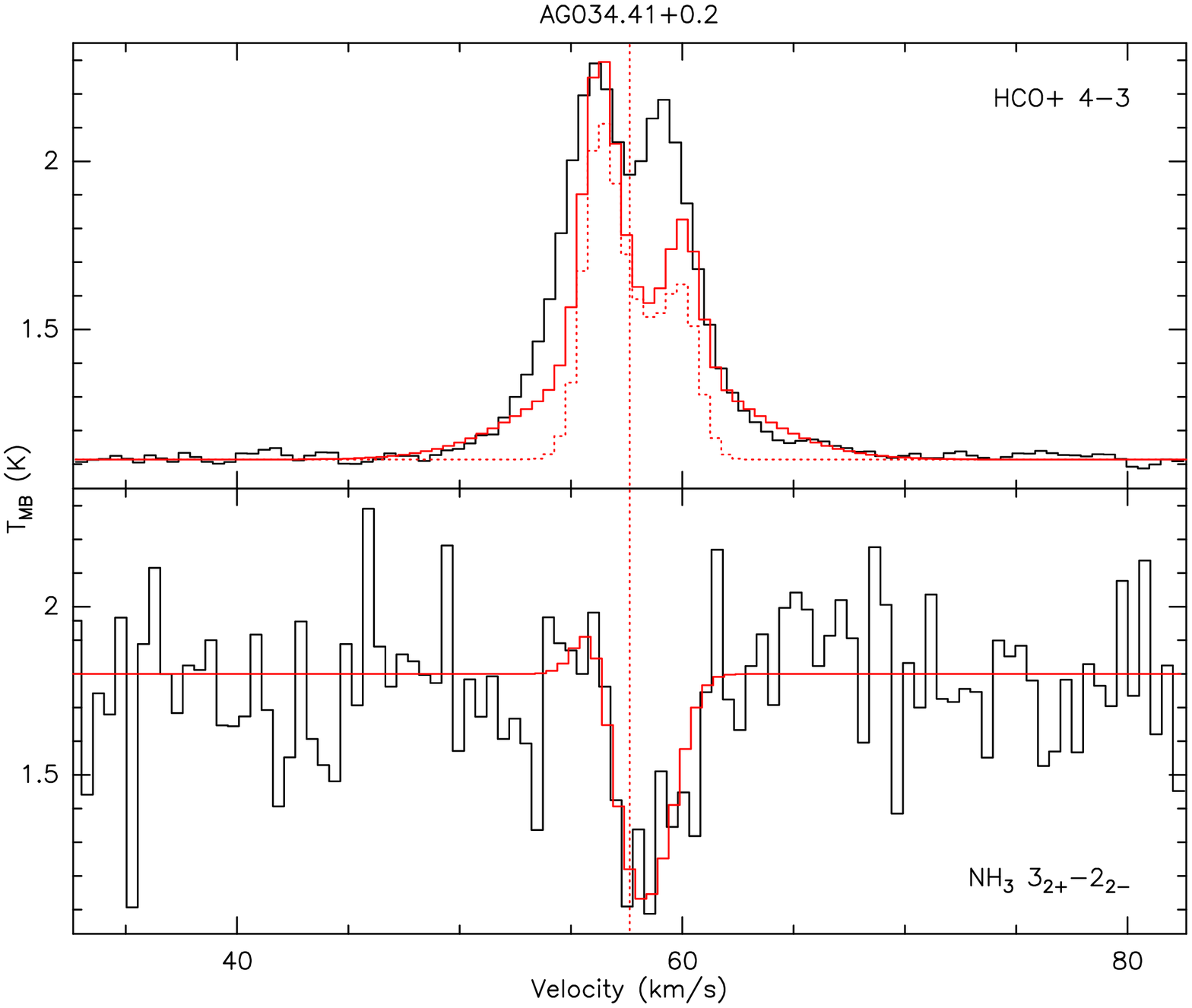}
}
\quad
\subfigure{
\includegraphics[width=0.35\textwidth,angle=0]{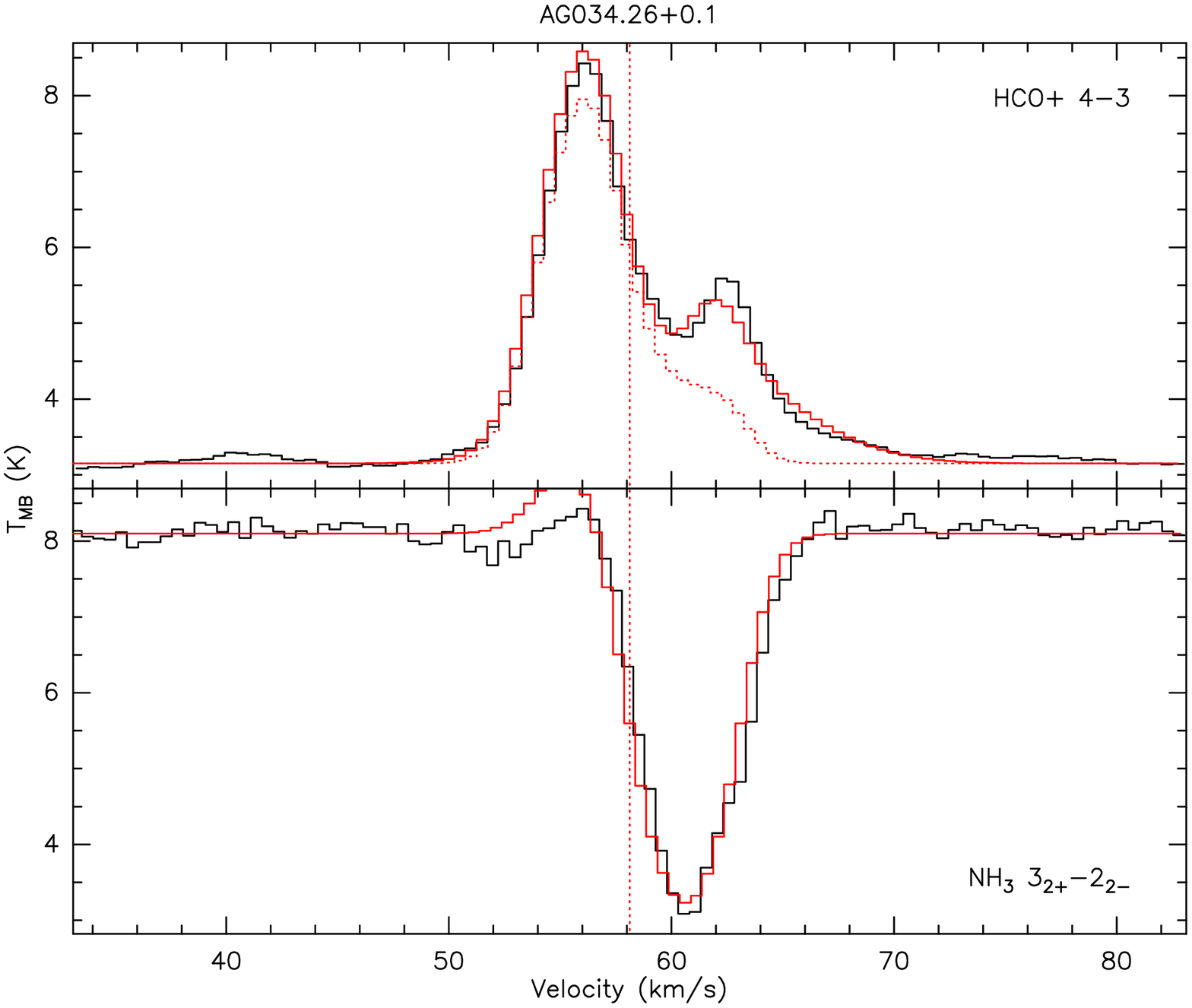}
}
\subfigure{
\includegraphics[width=0.35\textwidth,angle=0]{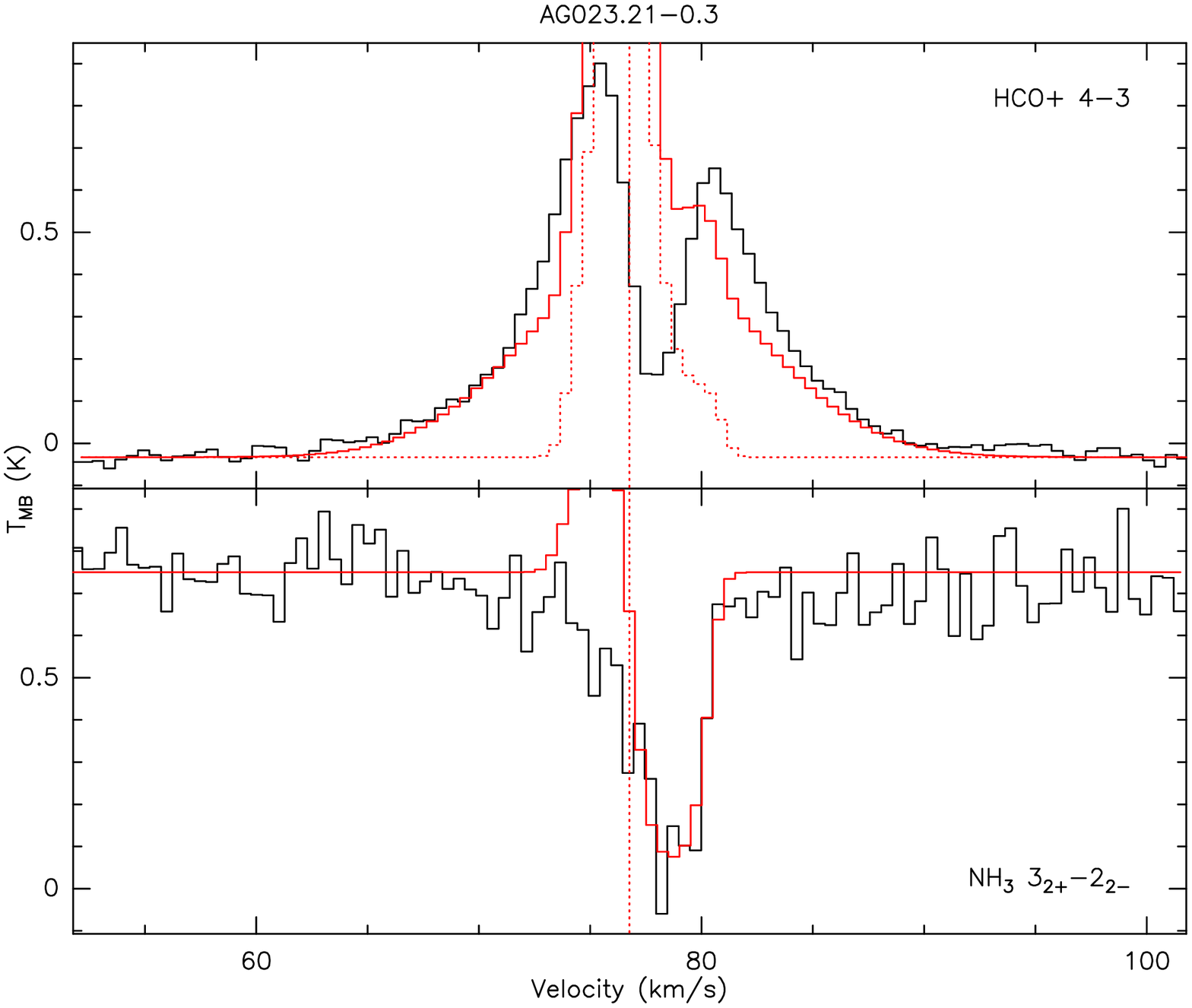}
}
\quad
\subfigure{
\includegraphics[width=0.35\textwidth,angle=0]{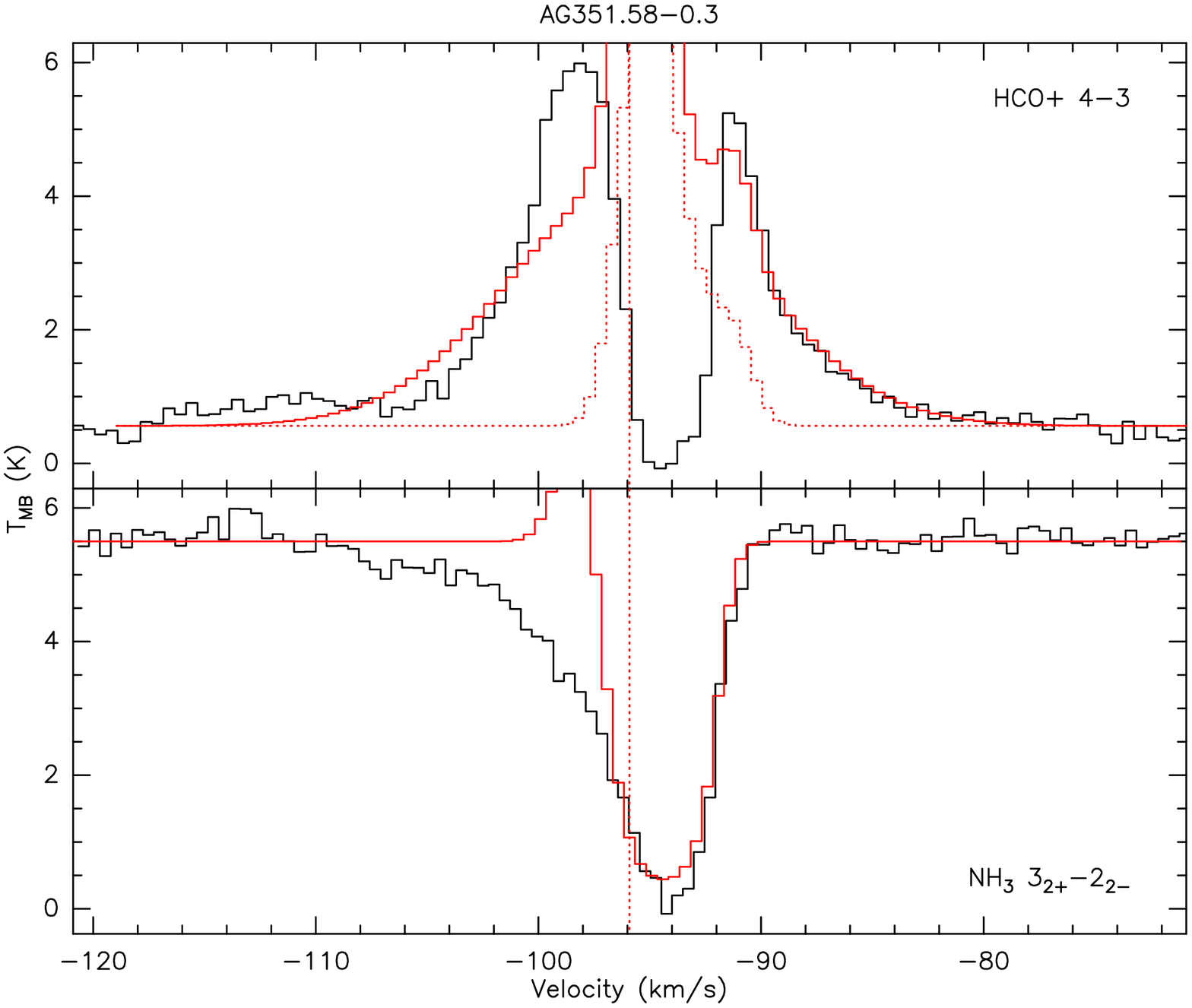}
}
\subfigure{
\includegraphics[width=0.35\textwidth,angle=0]{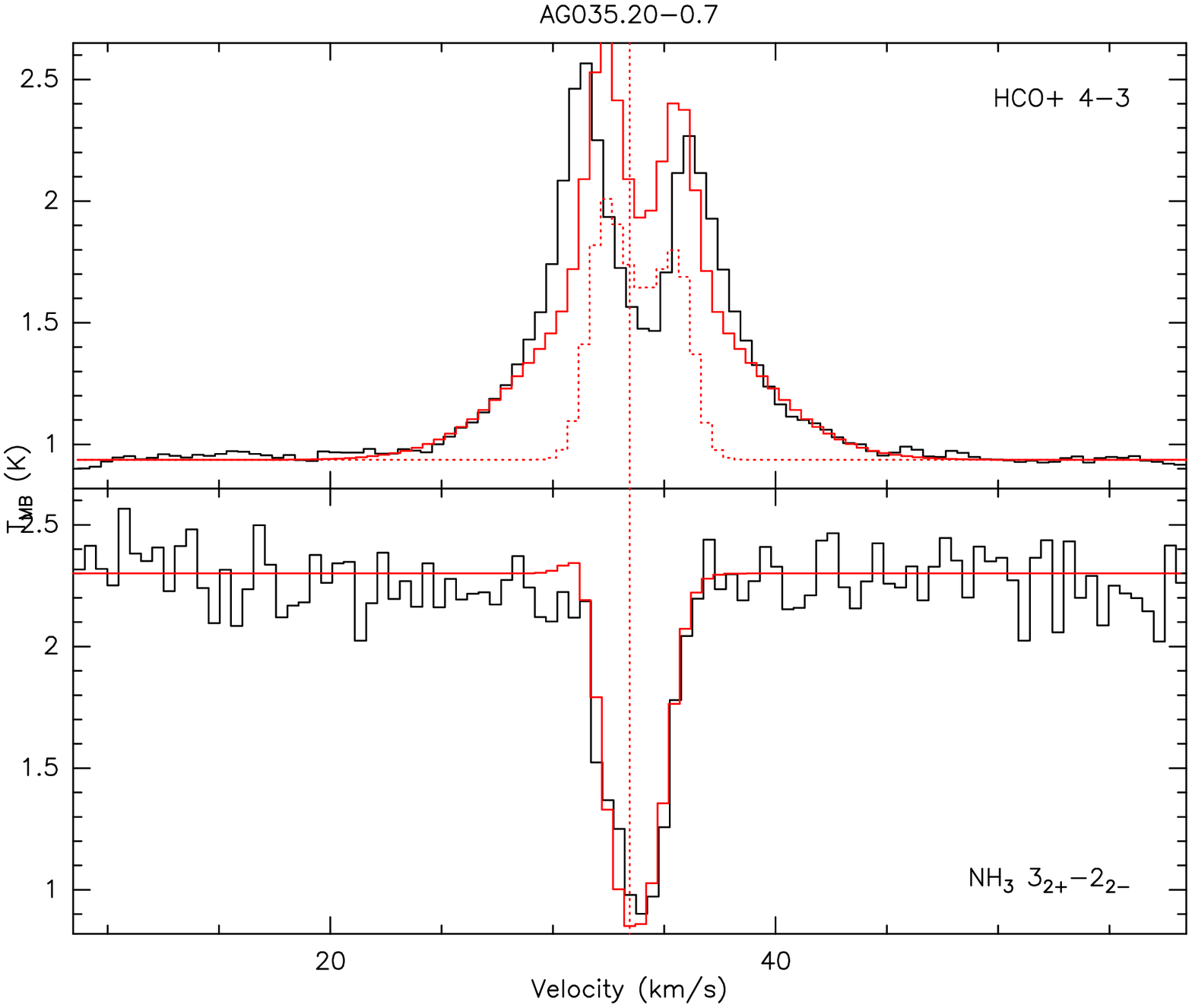}
}
\quad
\subfigure{
\includegraphics[width=0.35\textwidth,angle=0]{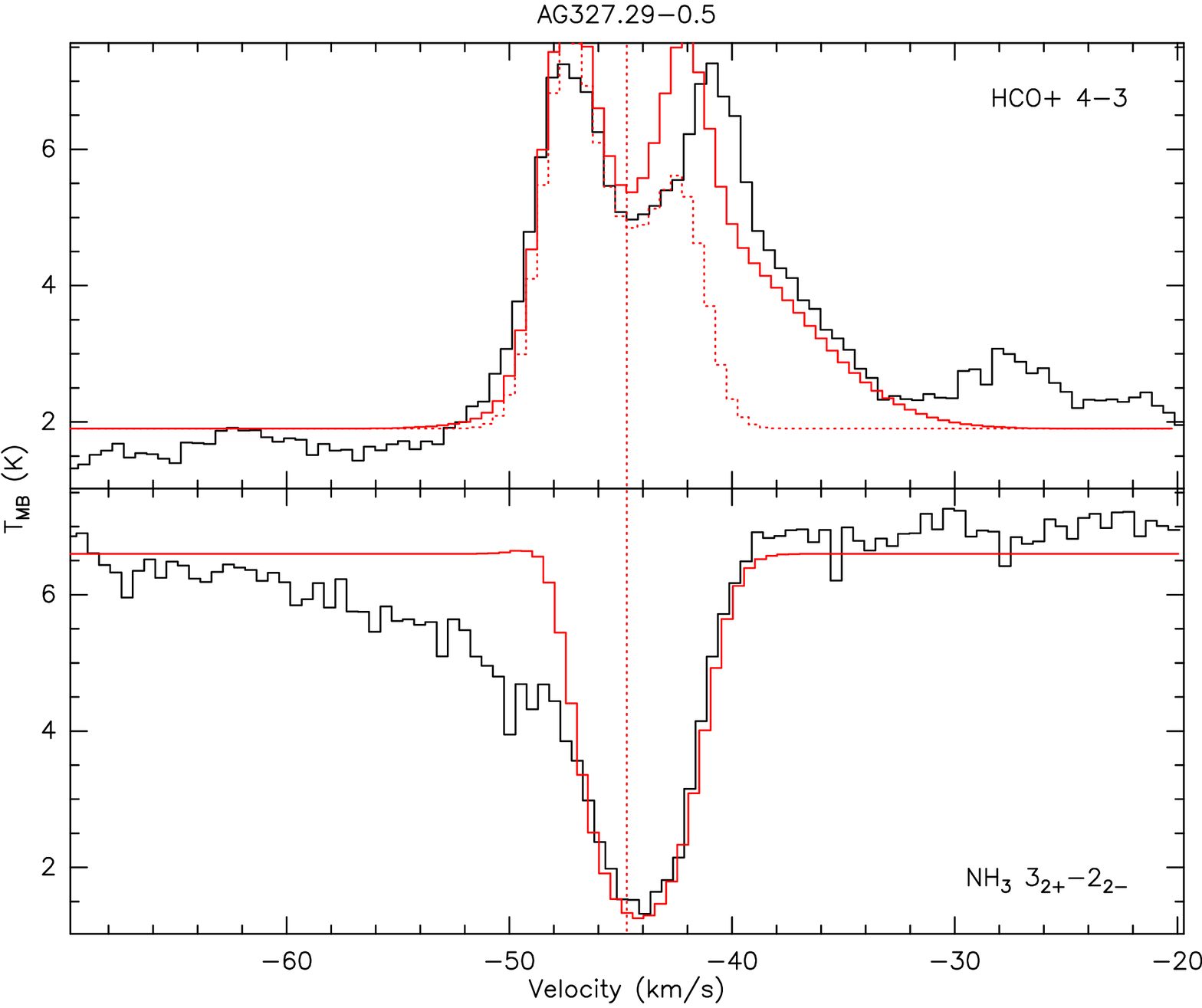}
}
\caption{\label{fig:hcop-nh3} Comparison of APEX \HCOP\ (4--3)
  observations with the SOFIA ammonia absorption in sources showing
  redshifted absorption. We show the RATRAN modeling result of \HCOP\
  (4--3) in red using variable abundances with and without (dotted)
  additional outflow components. }
\end{center}
\end{figure*}

Even more indications of an additional outflow component are seen in
most \HCOP\ lines. The dotted fits to the HCO$^+$ lines in
Fig.~\ref{fig:hcop-nh3} show the infall models with only an
adjustment of the HCO$^+$ abundance. In general, they cannot
reproduce the observed profile signatures, that is, the broad emission,
the deep self-absorption, and the relative intensities of the blue and
red peaks.  We therefore added  outflow emission to the RATRAN models of \HCOP\
in the
center plane of the envelope before the ray-tracing
step, as was done in \citet{mottram+2013}.  Examples of profiles that  can be fit satisfyingly with this
additional component are G34.26+0.1 and
G35.20-0.7, as shown in Fig.~\ref{fig:hcop-nh3}. The outflow components
have typical widths of 10~km/s and are in some cases slightly shifted from
the systemic velocities to account for the asymmetry of the outflow
lobes. For several clumps, the HCO$^+$ abundances are relatively small
but consistent with typical depletion factors found in
\citet{giannetti+2014}. Still, several HCO$^+$ profiles cannot be
reproduced with the simplicity of the spherical models, pointing to inherent difficulties in the interpretation of HCO$^+$ profiles as an
infall indicator.



\section{Discussion and  conclusions}

\begin{figure}[ht]
\begin{center}
\includegraphics[width=0.35\textwidth,angle=0]{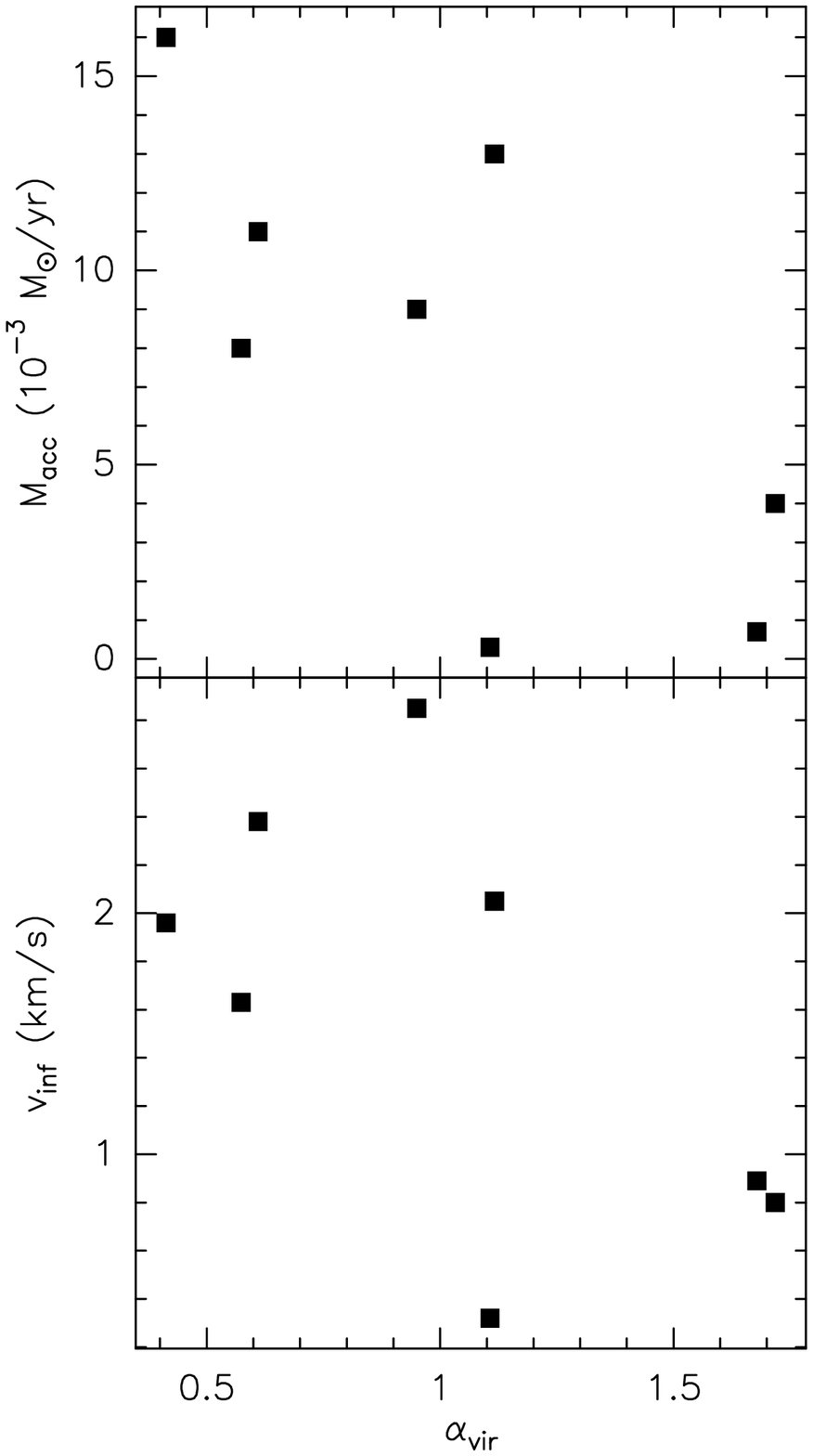}
\caption{\label{fig:mvir} Comparison of the infall velocities and
rates from the ammonia absorption measurements with the virial parameter
of the clumps from \citet{giannetti+2014}.}
\end{center}
\end{figure}

Adding the results from \citet{wyrowski+2012}, eight out of eleven
massive clumps have been found with redshifted absorption that
is indicative
of infall motions. This fraction of 72\% is substantially higher than
found in past searches for the blue-skewed profile signature, see, for example,
\citet{fuller+2005}, \citet{wu+2007}, \citet{lopez-sepulcre+2010}, \citet{reiter+2011}
and \citet{rygl+2013}.

From our comparison of ammonia absorption with line profiles of dense
gas tracers lines, we conclude that outflows are the most likely reason for hiding
infall motions in the clump envelopes. This effect is less prominent
in the THz ammonia spectra. While we see  the red lobe due to the high excitation even in emission in the extreme outflow
source
G5.89-0.4,
other sources are, if at all, only affected by blueshifted absorption
from the outflow in front of the continuum. These absorption wings
are similar to the blue emission wings in the CS (7--6) spectra, but do
not affect the measurements of the main redshifted absorption from
the envelope. The excitation in most outflows is too low to excite the
high critical density THz ammonia line ($n_{cr}(25\,{\rm K})
  \approx 10^9/\tau$~\percc), but are high enough for the submillimeter
lines, where already little outflowing gas strongly influences the profiles. 
The best agreement is found with HCO$^+$ (4--3), that is, for a line
with higher critical density than the (1--0) line and of a molecule
less enhanced in outflows than for instance CS and HCN \citep[e.g.,][and
  see line wings in Fig.~\ref{fig:densegas-g23}]{bachiller+2001}. 
The
simulations of \citet{smith+2013} instead predicted the (1--0) to be
the better probe of infall signatures, but these simulations did not
yet include any outflow activity.

From the RATRAN modeling of the absorption lines the infall
rates can be also derived as $\dot{M}=4 \pi R^2 m_{\rm H_2} n v$ assuming
spherical infall.  We list in Table 5 infall rates in the range
from 0.3 to 16~$10^{-3}\,M_\odot/$yr using the beam size
of the observations as radius. \citet{tan+2014} reported 14.6~$10^{-3}\,M_\odot/$yr
as a
characteristic infall rate for a free-falling clump of
1000~$M_\odot$ , which is within a factor of two
  comparable to our infall rate results considering the masses, column
  densities, and fractions of the free-fall velocity of 0.03 to 0.3 of
  our sources. We compare these infall rates in Fig. 7 to the virial parameters $\alpha_{\rm vir}=M_{\rm vir}/M$ of
the clumps derived by \citet{giannetti+2014} using \CSEO. The clumps
W43 and G31.41 from \citet{wyrowski+2012} were included, and the
dependence between virial parameter and the infall velocities is also
shown. There is a slight trend seen toward higher infall rates and
velocities for lower virial parameter values $\alpha_{\rm vir}$ 
  with Spearman correlation coefficients of $-0.55$ and $-0.43$,
  respectively. This is consistent with the expectation that the more
gravity dominates, the higher the inward accelerations become.
  It is likewise expected that the correlation of the infall rates with the mass of
  the clumps is even higher, with a Spearman coefficient of 0.86.

  The new observations show that infall on clump scales is ubiquitous
  throughout a wide range of evolutionary stages, including clumps in
  infrared dark clouds, from $L/M$ covering about $10-70$.  While we did
  not find a correlation of the infall rate or velocity with $L/M$
  (Spearman coefficients of $-0.29$ and $-0.14$, respectively), it is
  notable that the three sources with blueshifted emission have the
  two highest and the fifth highest $L/M$, either indicating that
  infall has stopped or that is has become undetectable as a
result of   relatively strong outflows and expanding motions compared to the
  colder envelope in which the other clumps show the infall signature.
  To search in more detail for evolutionary effects requires a larger sample
  and expanding this study to lower $L/M$. For this purpose, the
  572~GHz ortho ground-state line of ammonia might be a future option
  for SOFIA to continue absorption studies to sources that are
too cold to have
  detectable continuum at 1.8~THz.

\acknowledgements{
  We thank the referee for comments and suggestions that have helped
  to improve this manuscript. We thank J. Mottram for making the
  RATRAN code with additional outflow component available. This work
  was partially carried out within the Collaborative Research Council
  956, subproject A6, funded by the Deutsche Forschungsgemeinschaft
  (DFG).  Based [in part] on observations made with the NASA/DLR
  Stratospheric Observatory for Infrared Astronomy. SOFIA Science
  Mission Operations are conducted jointly by the Universities Space
  Research Association, Inc., under NASA contract NAS2-97001, and the
  Deutsches SOFIA Institut under DLR contract 50 OK 0901.
}

\bibliographystyle{aa}
\bibliography{../papers-all}

\begin{thebibliography}{}
\expandafter\ifx\csname natexlab\endcsname\relax\def\natexlab#1{#1}\fi

\bibitem[{{Bachiller} {et~al.}(2001){Bachiller}, {Perez Gutierrez}, {Kumar}, \&
  {Tafalla}}]{bachiller+2001}
{Bachiller}, M., {Perez Gutierrez}, M., {Kumar}, M., \& {Tafalla}, M. 2001,
  \aap, 372, 899

\bibitem[{{Beltr{\'a}n} {et~al.}(2006){Beltr{\'a}n}, {Cesaroni}, {Codella},
  {Testi}, {Furuya}, \& {Olmi}}]{beltran+2006}
{Beltr{\'a}n}, M.~T., {Cesaroni}, R., {Codella}, C., {et~al.} 2006, \nat, 443,
  427

\bibitem[{{Carey} {et~al.}(2009){Carey}, {Noriega-Crespo}, {Mizuno}, {Shenoy},
  {Paladini}, {Kraemer}, {Price}, {Flagey}, {Ryan}, {Ingalls}, {Kuchar},
  {Pinheiro Gon{\c c}alves}, {Indebetouw}, {Billot}, {Marleau}, {Padgett},
  {Rebull}, {Bressert}, {Ali}, {Molinari}, {Martin}, {Berriman}, {Boulanger},
  {Latter}, {Miville-Deschenes}, {Shipman}, \& {Testi}}]{carey+2009}
{Carey}, S.~J., {Noriega-Crespo}, A., {Mizuno}, D.~R., {et~al.} 2009, \pasp,
  121, 76

\bibitem[{{Churchwell} {et~al.}(2009){Churchwell}, {Babler}, {Meade},
  {Whitney}, {Benjamin}, {Indebetouw}, {Cyganowski}, {Robitaille}, {Povich},
  {Watson}, \& {Bracker}}]{churchwell+2009}
{Churchwell}, E., {Babler}, B.~L., {Meade}, M.~R., {et~al.} 2009, \pasp, 121,
  213

\bibitem[{{Evans}(2003)}]{evans2002}
{Evans}, II, N. 2003, in SFChem 2002: Chemistry as a Diagnostic of Star
  Formation, ed. {C.~L.~Curry \& M.~Fich}, 157

\bibitem[{{Fuller} {et~al.}(2005){Fuller}, {Williams}, \&
  {Sridharan}}]{fuller+2005}
{Fuller}, G.~A., {Williams}, S.~J., \& {Sridharan}, T.~K. 2005, \aap, 442, 949

\bibitem[{{Giannetti} {et~al.}(2014){Giannetti}, {Wyrowski}, {Brand},
  {Csengeri}, {Fontani}, {Walmsley}, {Nguyen Luong}, {Beuther}, {Schuller},
  {G{\"u}sten}, \& {Menten}}]{giannetti+2014}
{Giannetti}, A., {Wyrowski}, F., {Brand}, J., {et~al.} 2014, \aap, 570, A65

\bibitem[{{Ginsburg} {et~al.}(2013){Ginsburg}, {Glenn}, {Rosolowsky},
  {Ellsworth-Bowers}, {Battersby}, {Dunham}, {Merello}, {Shirley}, {Bally},
  {Evans}, {Stringfellow}, \& {Aguirre}}]{ginsburg+2013}
{Ginsburg}, A., {Glenn}, J., {Rosolowsky}, E., {et~al.} 2013, \apjs, 208, 14

\bibitem[{{Heyminck} {et~al.}(2012){Heyminck}, {Graf}, {G{\"u}sten}, {Stutzki},
  {H{\"u}bers}, \& {Hartogh}}]{heyminck+2012}
{Heyminck}, S., {Graf}, U.~U., {G{\"u}sten}, R., {et~al.} 2012, \aap, 542, L1

\bibitem[{{Hogerheijde} \& {van der Tak}(2000)}]{hogerheijde+2000}
{Hogerheijde}, M.~R. \& {van der Tak}, F.~F.~S. 2000, \aap, 362, 697

\bibitem[{{Klein} {et~al.}(2012){Klein}, {Hochg{\"u}rtel}, {Kr{\"a}mer},
  {Bell}, {Meyer}, \& {G{\"u}sten}}]{Klein+2012}
{Klein}, B., {Hochg{\"u}rtel}, S., {Kr{\"a}mer}, I., {et~al.} 2012, \aap, 542,
  L3

\bibitem[{{Kurayama} {et~al.}(2011){Kurayama}, {Nakagawa}, {Sawada-Satoh},
  {Sato}, {Honma}, {Sunada}, {Hirota}, \& {Imai}}]{kurayama+2011}
{Kurayama}, T., {Nakagawa}, A., {Sawada-Satoh}, S., {et~al.} 2011, \pasj, 63,
  513

\bibitem[{Leurini {et~al.}(2015)Leurini, Wyrowski, Wiesemeyer, Gusdorf,
  Guesten, Menten, Gerin, Levrier, Huebers, Jacobs, Ricken, \&
  Richter}]{leurini+2015}
Leurini, S., Wyrowski, F., Wiesemeyer, H., {et~al.} 2015, arXiv:1510.00366

\bibitem[{{L{\'o}pez-Sepulcre} {et~al.}(2010){L{\'o}pez-Sepulcre}, {Cesaroni},
  \& {Walmsley}}]{lopez-sepulcre+2010}
{L{\'o}pez-Sepulcre}, A., {Cesaroni}, R., \& {Walmsley}, C.~M. 2010, \aap, 517,
  A66

\bibitem[{{Marseille} {et~al.}(2010){Marseille}, {van der Tak}, {Herpin}, \&
  {Jacq}}]{marseille+2010}
{Marseille}, M.~G., {van der Tak}, F.~F.~S., {Herpin}, F., \& {Jacq}, T. 2010,
  \aap, 522, A40

\bibitem[{{Molinari} {et~al.}(2008){Molinari}, {Pezzuto}, {Cesaroni}, {Brand},
  {Faustini}, \& {Testi}}]{molinari+2008}
{Molinari}, S., {Pezzuto}, S., {Cesaroni}, R., {et~al.} 2008, \aap, 481, 345

\bibitem[{{Molinari} {et~al.}(2010){Molinari}, {Swinyard}, {Bally}, {Barlow},
  {Bernard}, {Martin}, {Moore}, {Noriega-Crespo}, {Plume}, {Testi}, {Zavagno},
  {Abergel}, {Ali}, {Andr{\'e}}, {Baluteau}, {Benedettini}, {Bern{\'e}},
  {Billot}, {Blommaert}, {Bontemps}, {Boulanger}, {Brand}, {Brunt}, {Burton},
  {Campeggio}, {Carey}, {Caselli}, {Cesaroni}, {Cernicharo}, {Chakrabarti},
  {Chrysostomou}, {Codella}, {Cohen}, {Compiegne}, {Davis}, {de Bernardis}, {de
  Gasperis}, {Di Francesco}, {di Giorgio}, {Elia}, {Faustini}, {Fischera},
  {Fukui}, {Fuller}, {Ganga}, {Garcia-Lario}, {Giard}, {Giardino}, {Glenn},
  {Goldsmith}, {Griffin}, {Hoare}, {Huang}, {Jiang}, {Joblin}, {Joncas},
  {Juvela}, {Kirk}, {Lagache}, {Li}, {Lim}, {Lord}, {Lucas}, {Maiolo},
  {Marengo}, {Marshall}, {Masi}, {Massi}, {Matsuura}, {Meny}, {Minier},
  {Miville-Desch{\^e}nes}, {Montier}, {Motte}, {M{\"u}ller}, {Natoli}, {Neves},
  {Olmi}, {Paladini}, {Paradis}, {Pestalozzi}, {Pezzuto}, {Piacentini},
  {Pomar{\`e}s}, {Popescu}, {Reach}, {Richer}, {Ristorcelli}, {Roy}, {Royer},
  {Russeil}, {Saraceno}, {Sauvage}, {Schilke}, {Schneider-Bontemps},
  {Schuller}, {Schultz}, {Shepherd}, {Sibthorpe}, {Smith}, {Smith},
  {Spinoglio}, {Stamatellos}, {Strafella}, {Stringfellow}, {Sturm}, {Taylor},
  {Thompson}, {Tuffs}, {Umana}, {Valenziano}, {Vavrek}, {Viti}, {Waelkens},
  {Ward-Thompson}, {White}, {Wyrowski}, {Yorke}, \& {Zhang}}]{molinari+2010b}
{Molinari}, S., {Swinyard}, B., {Bally}, J., {et~al.} 2010, \pasp, 122, 314

\bibitem[{{Mottram} {et~al.}(2013){Mottram}, {van Dishoeck}, {Schmalzl},
  {Kristensen}, {Visser}, {Hogerheijde}, \& {Bruderer}}]{mottram+2013}
{Mottram}, J.~C., {van Dishoeck}, E.~F., {Schmalzl}, M., {et~al.} 2013, \aap,
  558, A126

\bibitem[{{Peng} {et~al.}(2010){Peng}, {Wyrowski}, {van der Tak}, {Menten}, \&
  {Walmsley}}]{peng+2010}
{Peng}, T.-C., {Wyrowski}, F., {van der Tak}, F.~F.~S., {Menten}, K.~M., \&
  {Walmsley}, C.~M. 2010, \aap, 520, A84

\bibitem[{{Reiter} {et~al.}(2011){Reiter}, {Shirley}, {Wu}, {Brogan},
  {Wootten}, \& {Tatematsu}}]{reiter+2011}
{Reiter}, M., {Shirley}, Y.~L., {Wu}, J., {et~al.} 2011, \apj, 740, 40

\bibitem[{{Rolffs} {et~al.}(2011){Rolffs}, {Schilke}, {Wyrowski}, {Menten},
  {G{\"u}sten}, \& {Bisschop}}]{rolffs+2011}
{Rolffs}, R., {Schilke}, P., {Wyrowski}, F., {et~al.} 2011, \aap, 527, A68

\bibitem[{{Rowan-Robinson}(1980)}]{rowan-robinson1980}
{Rowan-Robinson}, M. 1980, \apjs, 44, 403

\bibitem[{{Rygl} {et~al.}(2013){Rygl}, {Wyrowski}, {Schuller}, \&
  {Menten}}]{rygl+2013}
{Rygl}, K.~L.~J., {Wyrowski}, F., {Schuller}, F., \& {Menten}, K.~M. 2013,
  \aap, 549, A5

\bibitem[{{Schuller} {et~al.}(2009){Schuller}, {Menten}, {Contreras},
  {Wyrowski}, {Schilke}, {Bronfman}, {Henning}, {Walmsley}, {Beuther},
  {Bontemps}, {Cesaroni}, {Deharveng}, {Garay}, {Herpin}, {Lefloch}, {Linz},
  {Mardones}, {Minier}, {Molinari}, {Motte}, {Nyman}, {Reveret}, {Risacher},
  {Russeil}, {Schneider}, {Testi}, {Troost}, {Vasyunina}, {Wienen}, {Zavagno},
  {Kovacs}, {Kreysa}, {Siringo}, \& {Wei{\ss}}}]{schuller+2009}
{Schuller}, F., {Menten}, K.~M., {Contreras}, Y., {et~al.} 2009, \aap, 504, 415

\bibitem[{{Smith} {et~al.}(2013){Smith}, {Shetty}, {Beuther}, {Klessen}, \&
  {Bonnell}}]{smith+2013}
{Smith}, R.~J., {Shetty}, R., {Beuther}, H., {Klessen}, R.~S., \& {Bonnell},
  I.~A. 2013, \apj, 771, 24

\bibitem[{{Sollins} {et~al.}(2005){Sollins}, {Zhang}, {Keto}, \&
  {Ho}}]{sollins+2005}
{Sollins}, P.~K., {Zhang}, Q., {Keto}, E., \& {Ho}, P.~T.~P. 2005, \apjl, 624,
  L49

\bibitem[{{Tan} {et~al.}(2014){Tan}, {Beltr{\'a}n}, {Caselli}, {Fontani},
  {Fuente}, {Krumholz}, {McKee}, \& {Stolte}}]{tan+2014}
{Tan}, J.~C., {Beltr{\'a}n}, M.~T., {Caselli}, P., {et~al.} 2014, Protostars
  and Planets VI, 149

\bibitem[{{van der Tak} {et~al.}(2013){van der Tak}, {Chavarr{\'{\i}}a},
  {Herpin}, {Wyrowski}, {Walmsley}, {van Dishoeck}, {Benz}, {Bergin},
  {Caselli}, {Hogerheijde}, {Johnstone}, {Kristensen}, {Liseau}, {Nisini}, \&
  {Tafalla}}]{vdtak+2013}
{van der Tak}, F.~F.~S., {Chavarr{\'{\i}}a}, L., {Herpin}, F., {et~al.} 2013,
  \aap, 554, A83

\bibitem[{{Wright} {et~al.}(2010){Wright}, {Eisenhardt}, {Mainzer}, {Ressler},
  {Cutri}, {Jarrett}, {Kirkpatrick}, {Padgett}, {McMillan}, {Skrutskie},
  {Stanford}, {Cohen}, {Walker}, {Mather}, {Leisawitz}, {Gautier}, {McLean},
  {Benford}, {Lonsdale}, {Blain}, {Mendez}, {Irace}, {Duval}, {Liu}, {Royer},
  {Heinrichsen}, {Howard}, {Shannon}, {Kendall}, {Walsh}, {Larsen}, {Cardon},
  {Schick}, {Schwalm}, {Abid}, {Fabinsky}, {Naes}, \& {Tsai}}]{wright+2010}
{Wright}, E.~L., {Eisenhardt}, P.~R.~M., {Mainzer}, A.~K., {et~al.} 2010, \aj,
  140, 1868

\bibitem[{{Wu} {et~al.}(2007){Wu}, {Henkel}, {Xue}, {Guan}, \&
  {Miller}}]{wu+2007}
{Wu}, Y., {Henkel}, C., {Xue}, R., {Guan}, X., \& {Miller}, M. 2007, \apjl,
  669, L37

\bibitem[{{Wyrowski} {et~al.}(2012){Wyrowski}, {G{\"u}sten}, {Menten},
  {Wiesemeyer}, \& {Klein}}]{wyrowski+2012}
{Wyrowski}, F., {G{\"u}sten}, R., {Menten}, K.~M., {Wiesemeyer}, H., \&
  {Klein}, B. 2012, \aap, 542, L15

\bibitem[{{Zhang} \& {Ho}(1997)}]{zhang+1997}
{Zhang}, Q. \& {Ho}, P.~T.~P. 1997, \apj, 488, 241

\end{thebibliography}

\appendix

\section{Spectra of dense gas tracers}

\begin{figure*}[ht]
\begin{center}
\includegraphics[height=0.9\textwidth,angle=0]{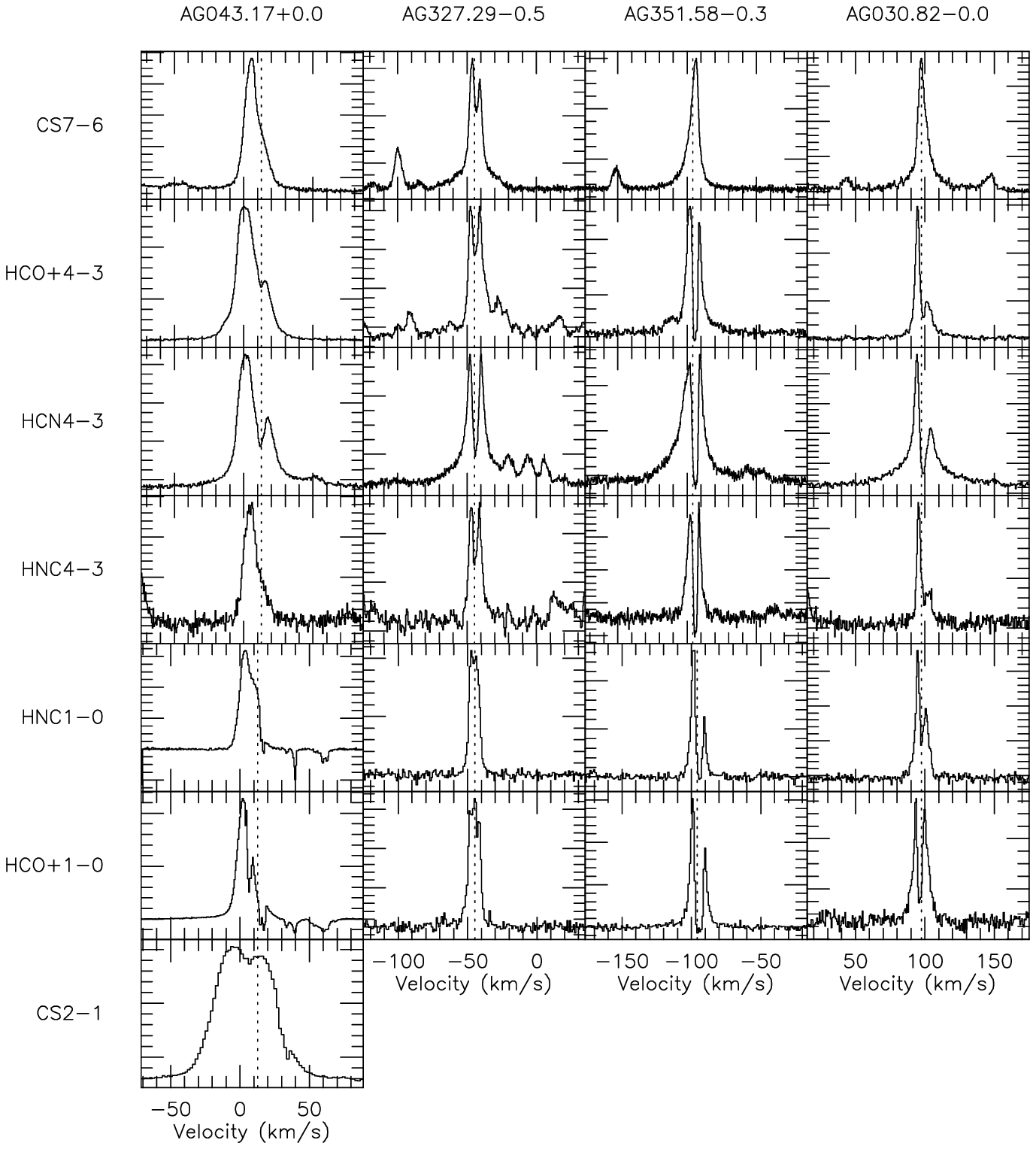}
\caption{\label{fig:densegas-1} Ground-based observations of
  millimeter and submillimeter transitions of the dense gas tracers
  HCN/HNC/CS/HCO$^+$. The systemic velocity from \CSEO\ (3--2) is
  indicated with a dashed line. }
\end{center}
\end{figure*}

\begin{figure*}[ht]
\begin{center}
\includegraphics[height=0.9\textwidth,angle=0]{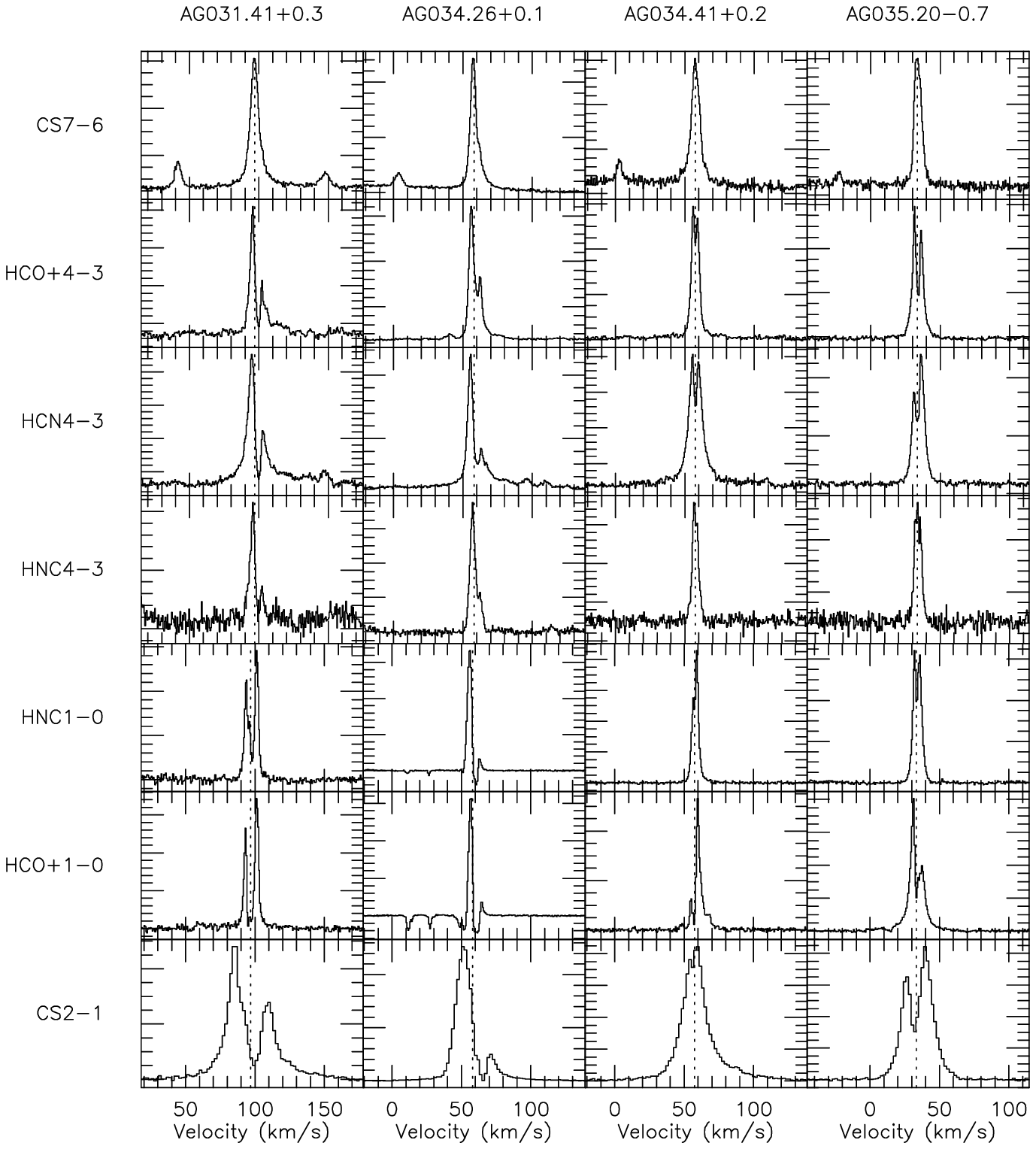}
\caption{\label{fig:densegas-2} Ground-based observations of
  millimeter and submillimeter transitions of the dense gas tracers
  HCN/HNC/CS/HCO$^+$. The systemic velocity from \CSEO\ (3--2) is
  indicated with a dashed line.}
\end{center}
\end{figure*}

\end{document}